\documentclass[%
reprint,
t,nobibnotes,
amsmath,amssymb,
prl,
]{revtex4-2}

\usepackage{graphicx}
\usepackage{dcolumn}
\usepackage{bm}
\usepackage{array}
\newcolumntype{M}[1]{>{\centering\arraybackslash}m{#1}}
\usepackage{caption}
\usepackage{subcaption}
\usepackage{float}
\usepackage{multirow}
\usepackage{comment}
\usepackage[dvipsnames]{xcolor}
\definecolor{firetruck}{RGB}{206,32,41}
\definecolor{sapphire}{RGB}{15,82,186}
\definecolor{myred}{RGB}{192, 0, 0}

\begin{document}

\preprint{APS/123-QED}

\title{An aeroacoustic mechanism to explain universal behavior in\\hypersonic wake flow oscillations}

\author{Premika S. Thasu}
 \email{Email address: premikat@iisc.ac.in}
\author{Gaurav Kumar}
 \thanks{presently at the University of Nevada, Reno}
\author{Subrahmanyam Duvvuri}
 \email{Corresponding author; email address: subrahmanyam@iisc.ac.in}

\affiliation{%
 Turbulent Shear Flow Physics and Engineering Laboratory\\
 Department of Aerospace Engineering, Indian Institute of Science, Bengaluru 560012, India\vspace*{2mm}
}%

\date{\today}

\begin{abstract}
Recent experimental studies reveal that the near-wake region of a circular cylinder at hypersonic Mach numbers exhibits self-sustained flow oscillations. The oscillation frequency was found to have a universal behavior. Experimental observations suggest an aeroacoustic feedback loop to be the driving mechanism of oscillations. An analytical aeroacoustic model which predicts the experimentally observed frequencies and explains the universal behavior is presented here. The model provides physical insights and informs of flow regimes where deviations from universal behavior are to be expected.
\end{abstract}

\maketitle
    

The wake of a 2D circular cylinder in the incompressible flow regime is a celebrated canonical problem in fluid dynamics \cite{roshko1961experiments,Wu1972Wakes,Williamson1996Vortex}. Historically, the cylinder wake problem was at the forefront of research in fluid dynamics, with motivation coming partly from practical issues in understanding and predicting hydrodynamic drag force on bluff bodies \cite{Wu1972Wakes,Williamson1996Vortex,Roshko1993BBaerodynamics}. The canonical cylinder wake contains many of the key elements of a generic bluff body wake. In the incompressible flow regime the physics of the wake is governed by a single non-dimensional parameter: the Reynolds number $Re_D = (\rho_{\infty}\, U_{\infty} D)/\mu_{\infty}$. Here $U_{\infty}$ is the freestream flow velocity, $D$ is the cylinder diameter, and $\rho_{\infty}$ and $\mu_{\infty}$ are the density and dynamic viscosity, respectively, of the freestream fluid.

\begin{figure}
    \centering
    \includegraphics[width=0.48\textwidth]{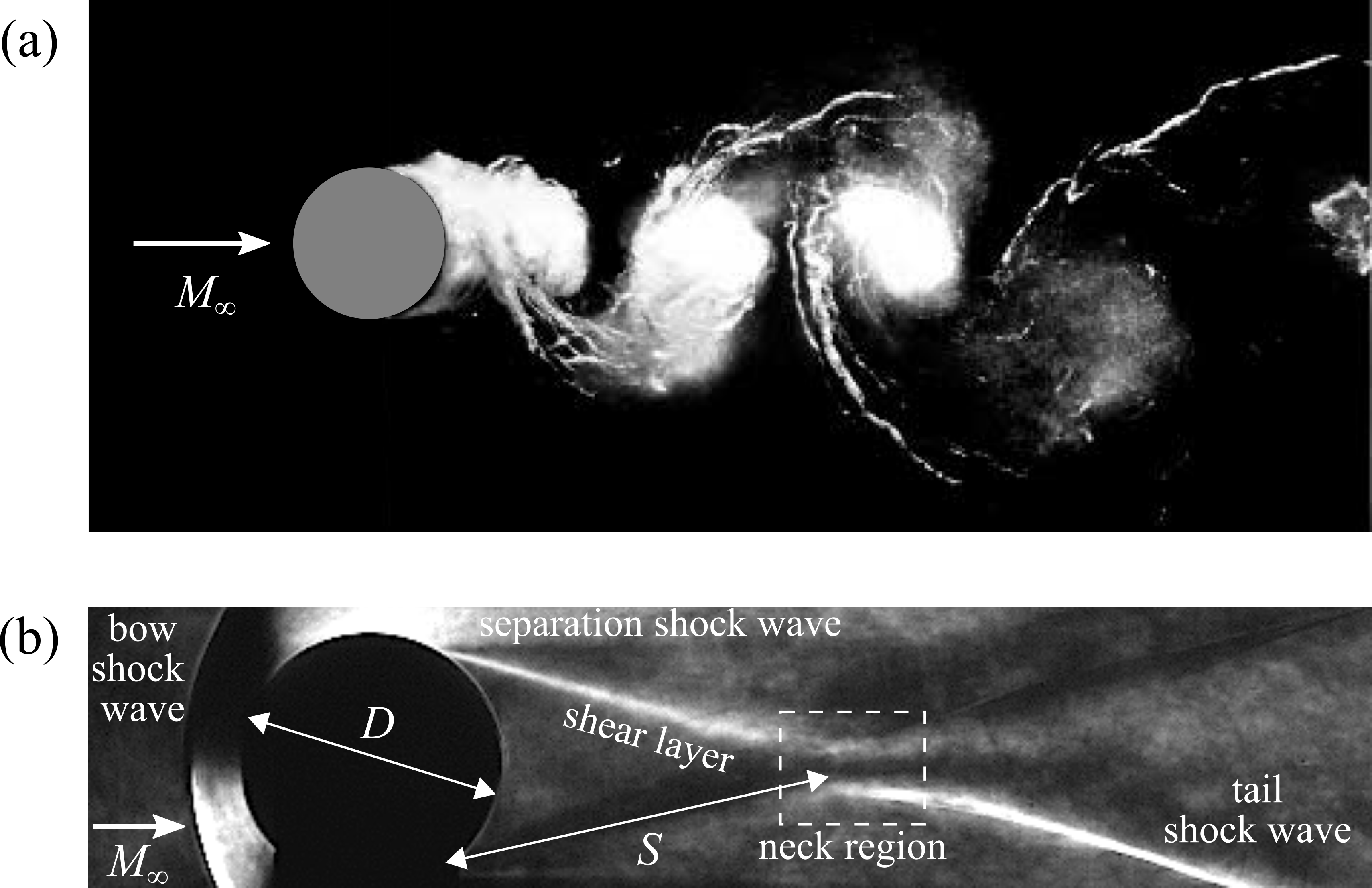}
    \caption{(a) An instantaneous snapshot of cylinder wake flow in water at $M_\infty = 0.01$ and $Re_D = 1.4 \times 10^5$ \cite{Djeridi2003NearWake}. The shedding vortices from the top and bottom of the cylinder aftbody, with opposing sense of rotation, are visualized by controlled cavitation \cite{Djeridi2003NearWake}.  (b) An instantaneous density gradient map around a circular cylinder at $M_\infty = 6$ and $Re_D = 2.8 \times 10^5$ obtained using the optical imaging technique of schlieren \cite{Thasu2022Strouhal}.}
    \label{fig:wake_comp}
\end{figure}

Periodic vortex shedding from the aftbody of the cylinder, which is a commonly observed flow feature in incompressible cylinder wakes, gives the flow a visually appealing character. (This is popularly known as the \emph{K\'{a}rm\'{a}n vortex street}; see Fig.~\ref{fig:wake_comp}a.) As $Re_D$ is increased starting from $Re_D \ll 1$, the phenomenon of vortex shedding manifests at $Re \approx 47$ \cite{Jackson1987ABodies}. Vortex shedding gives the wake a characteristic timescale, which is written in the form of a non-dimensional frequency referred to as the Strouhal number $St = (f\,L)/U$. Here $f$ is the characteristic frequency of flow oscillations in the wake region, and $L$ and $U$ are characteristic length and velocity scales, respectively. It is noted that $St$ is a function of $Re_D$. An interesting and fundamentally important aspect of $St$ is its universal behavior; when the shedding frequency is scaled using the wake width and a characteristic wake velocity, $St$ attains invariance for $Re_{D} > 300$, with a value of approximately 0.164 \cite{Roshko1952OnStreets, Roshko1954OnBodies,Bearman1967vortex,Griffin1981universal,Williamson1996Vortex}. This universal behaviour holds across a broad range of Reynolds numbers for cylinders and also for other bluff body shapes in the incompressible flow regime \cite{Griffin1981universal,Awbi1981assessment,anderson1996universal}.

Relaxing the incompressibility condition presents a more general problem of the wake flow in the compressible regime, where the Mach number becomes an additional (and independent) governing parameter along with the Reynolds number. The freestream flow Mach number is defined as $M_\infty = U_{\infty}/a_{\infty}$, where $a_{\infty}$ is the acoustic wave speed. The 2D cylinder wake at supersonic and hypersonic Mach numbers, where the flow is compressible, is very distinct from its incompressible flow regime counterpart. As a representative example, Fig.~\ref{fig:wake_comp}b shows an instantaneous flow density gradient map for a cylinder at $M_\infty = 6$. Distinction between incompressible and compressible cylinder wake flows are qualitatively very evident in Fig.~\ref{fig:wake_comp}. For $M_{\infty} > 1$, shock waves appear in the flow and the vortex shedding phenomenon disappears. The near-wake region is characterized by the formation of two shear layers, which are symmetric about the cylinder centerline.  

The canonical compressible cylinder wake problem has received much less scientific attention as compared to the incompressible problem. The near-wake periodic flow unsteadiness in the incompressible regime has been the subject of several detailed studies \cite{Wu1972Wakes,Williamson1996Vortex}, which span a period of over a hundred years \cite{Raghu2005vortexstreet}, and the unsteadiness mechanisms are reasonably well understood \cite{Abernathy1962Formation,Gerrard1966Mechanics,Perry1982Vortex,Williamson1996Vortex}. Whereas for the supersonic/hypersonic flow regime, it is only within the past decade that the near-wake flow region was discovered to exhibit coherent and periodic oscillations \cite{Schmidt2015OscillationsMach4,Thasu2022Strouhal,Awasthi2022SupersonicDynamics} \footnote{A schlieren video of these flow oscillations is available as supplementary material to reference \cite{Thasu2022Strouhal}.}. The oscillations were found to have a single characteristic frequency. Interestingly, the oscillation Strouhal number, formed using the shear layer length and freestream velocity, exhibits universal behavior. At high-supersonic and hypersonic Mach numbers and across a range of Reynolds numbers, the Strouhal number was found to be invariant, taking a value of approximately 0.48 \cite{Schmidt2015OscillationsMach4,Thasu2022Strouhal}. 

The nature of cylinder wake unsteadiness in the compressible flow regime is fundamentally very different from the incompressible flow scenario. Driven by experimental observations, literature proposes the following hypothesis: an aeroacoustic feedback mechanism in the near-wake causes and sustains flow oscillations \cite{Schmidt2015OscillationsMach4,Thasu2022Strouhal}. Based on this hypothesis, here we develop a quantitative aeroacoustic model with no empiricism to explain oscillations observed in the supersonic/hypersonic cylinder wake. The model successfully predicts the oscillation frequencies reported from experiments, and thereby provides a clear physical understanding of the phenomenon. Further, the model also explains the experimentally observed universal behavior and informs of flow regimes where deviations from universal behavior are to be expected. Table~\ref{tab:exp_data} summarizes the experimental data available in literature for near-wake oscillations at high Mach numbers. $S$ is the shear layer length (see Fig.~\ref{fig:wake_comp}b), and two Strouhal numbers are defined as $St_{D} = (f\,D)/U_\infty$ and $St_{S} = (f\,S)/U_\infty$. Data from Table~\ref{tab:exp_data} are used for the present model development exercise and for validation of model predictions.
\begin{table}
    \centering
    \def\arraystretch{1.3}%
    \begin{tabular}{|c|c|c|c|c|c|}
        \hline
        & & & & & \\[-1mm]
        & $\bm{M_\infty}$ & $\bm{Re_D}\left(\times 10^4\right)$ & $\bm{\frac{S}{D}}$ & $\bm{St_D}$ & $\bm{St_S}$ \\
        & & & & & \\[-1mm]
        \hline
        \multirow{8}{6em}{\citet{Schmidt2015OscillationsMach4}} & \multirow{8}{2em}{\centering 4} & 2 & 1.52 & 0.299 & 0.45 \\
        & & 4 & 1.84 & 0.308 & 0.56 \\
        & & 5 & 1.23 & 0.339 & 0.42 \\
        & & 9 & 1.28 & 0.368 & 0.47 \\
        & & 13 & 1.35 & 0.398 & 0.53 \\
        & & 21 & 1.27 & 0.459 & 0.58 \\
        & & 29 & 1.07 & 0.448 & 0.48 \\
        & & 49 & 0.97 & 0.478 & 0.46 \\
        \hline
        \multirow{6}{6em}{\citet{Thasu2022Strouhal}} & \multirow{6}{2em}{\centering 6} & 23 & 1.49 & 0.341 & 0.5 \\
        & & 28 & 1.39 & 0.346 & 0.48 \\
        & & 30 & 1.38 & 0.349 & 0.48 \\
        & & 40 & 1.38 & 0.355 & 0.49 \\
        & & 43 & 1.3 & 0.361 & 0.47 \\
        & & 50 & 1.27 & 0.372 & 0.47 \\
        \hline
    \end{tabular}
    \caption{Strouhal number data from experiments.}
    \label{tab:exp_data}
\end{table}

\begin{figure}
    \centering
    \includegraphics[width=0.48\textwidth]{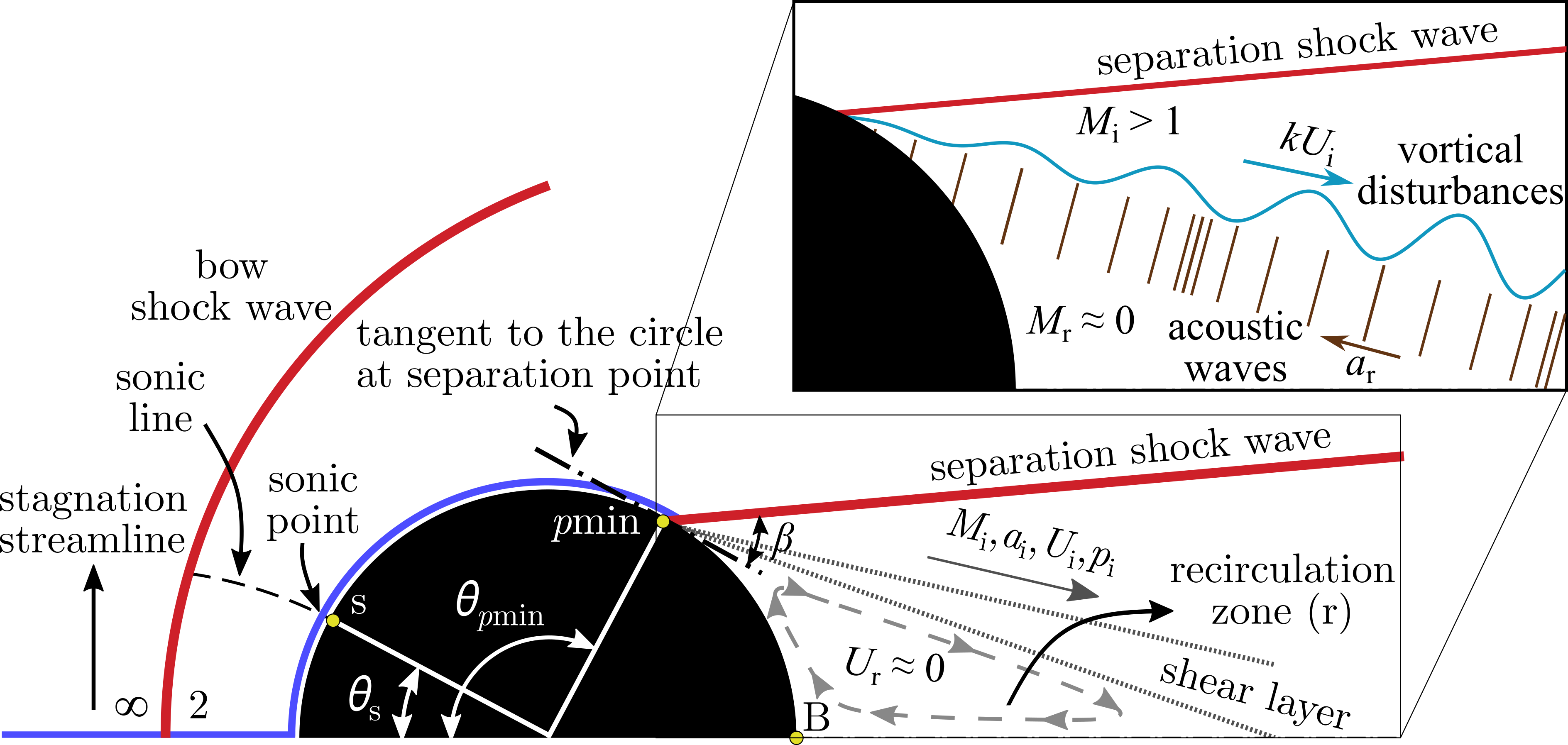}
    \caption{A schematic illustration of the flow structure over the top half of a supersonic/hypersonic cylinder.}
    \label{fig:cylWake_schem}
\end{figure}

We begin with a brief description of the key flow features (see Fig.~\ref{fig:cylWake_schem}). Given the flow symmetry about the cylinder centerline, only the top half of the cylinder and flow are depicted in the figure. A steady bow shock wave forms upstream of the cylinder, and the flow downstream of the shock wave in region 2 is subsonic. The subsonic fluid accelerates as it moves around the cylinder, attains Mach 1 at the sonic line, and further accelerates to supersonic Mach numbers as the flow expands around the cylinder. Further downstream the flow separates from the cylinder surface (due to the limitation on the maximum turn angle of supersonic flows) and generates a separation shock wave. Flow separation on the top and bottom surfaces of the cylinder results in the formation of symmetric supersonic shear layers on either side of the centerline. The region of intersection of the two shear layers is referred to as the ``neck'' of the wake (marked in Fig.~\ref{fig:wake_comp}b). The shear layers and the cylinder surface enclose two regions of subsonic recirculating flow with opposing sense of rotation. Downstream of the neck the flow turns parallel to the freestream through tail shock waves that are generated at the neck region.

The physical picture of flow oscillations that we build on is the following: interaction between the two shear layers in the neck region leads to an aeroacoustic feedback loop and sustains periodic flow oscillations. The inset in Fig.~\ref{fig:cylWake_schem} illustrates this mechanism, which comprises of four distinct phases:
\vspace*{-2mm}\begin{enumerate}
    \item downstream propagation and amplification of vortical disturbances (generated by flow instabilities) in the shear layers;
    \vspace*{-2mm}\item scattering, or generation of acoustic disturbances, at the neck region due to interaction between the two unsteady shear layers;
    \vspace*{-2mm}\item upstream propagation of acoustic waves along the subsonic portion of the shear layers;
    \vspace*{-2mm}\item receptivity of the shear layers to the acoustic waves, resulting in the excitation of vortical disturbances in their upstream regions.
\end{enumerate}\vspace*{-2mm}
It is noted that a broadly similar mechanism to the one outlined above is at play when air flows at high subsonic or supersonic speeds over open cavities, and leads to periodic unsteadiness and emission of acoustic tones. Some insights from open cavity flow literature, particularly the modeling framework used therein \cite{Powell1953edgetones, Powell1964VortexSound, Rossiter1964Wind-tunnelSpeeds}, are leveraged for the present effort.

By considering the feedback loop to be linear, and matching the wavespeed to wavelength ratio (\emph{i.e.}, frequency) between downstream-propagating vortical disturbances and upstream-propagating acoustic disturbances, the following expression can be obtained for the disturbance frequency $f$ \cite{Powell1953edgetones, Powell1964VortexSound, Rossiter1964Wind-tunnelSpeeds}: 
\begin{equation}\label{eq:ross_original}
    f = \left(\frac{1}{S}\right)\left(\frac{m-\phi}{\frac{1}{a_\mathrm{r}}+\frac{1}{kU_\mathrm{i}}}\right).
\end{equation}
Here $U_\mathrm{i}$ is the flow velocity downstream of the separation shock wave (see Fig.~\ref{fig:cylWake_schem}),  $kU_\mathrm{i}$ is the propagation speed of vortical disturbances (with $k$ being a constant), $a_\mathrm{r}$ is the speed of the acoustic waves that propagate upstream along the shear layer of length $S$, integer $m$ is the mode number for the oscillations, and $\phi$ is the phase difference between the vortical and acoustic disturbances at the neck region. The flow oscillation timescale is taken to be the same as the vortical and acoustic disturbance timescale, \emph{i.e.}, $f$ also denotes the wake oscillation frequency. The Strouhal number $St_D$ of wake oscillations can then be written as
\begin{equation}\label{eq:ross_StD}
    St_D = \frac{fD}{U_\infty} = \left(\frac{D}{S}\right)\left(\frac{m-\phi}{\frac{U_\infty}{a_\mathrm{r}}+\frac{U_\infty}{kU_\mathrm{i}}}\right).
\end{equation}

Obtaining the exact conditions downstream of the separation shock waves requires detailed flow computations. However, with certain simplifications, the flow conditions can be estimated reasonably well without resorting to computations. An acceptable modeling approximation is to consider flow along the stagnation streamline (marked in Fig.~\ref{fig:cylWake_schem}) downstream of the bow shock wave to be isentropic. Flow stagnation properties downstream of the bow shock wave (region 2 in Fig.~\ref{fig:cylWake_schem}) are obtained by assuming the bow shock wave to be locally normal in the region close to the cylinder centerline \cite{Liepmann1957ElementsDynamics}. The pressure minimum location on the cylinder surface, marked as $\theta_{p\mathrm{min}}$ in Fig.~\ref{fig:cylWake_schem}, occurs slightly upstream of the region where the separation shock wave forms \cite{Hinman2017ReynoldsWakes}. Flow properties at $\theta_{p\mathrm{min}}$ are obtained using the Prandtl-Meyer expansion fan theory \cite{Liepmann1957ElementsDynamics} with the flow turn angle given by the difference in angle between $\theta_{p\mathrm{min}}$ \cite{Hinman2017ReynoldsWakes} and the sonic point location $\theta_\mathrm{s}$ \cite{Sinclair2017ACylinder}. The solution of the separation shock wave requires information on at least one flow property downstream. From earlier studies of compressible cylinder wake flows aimed at understading the mean (time-averaged) flow structure \cite{McCarthy1964A5.7,Dewey1965NearSpeeds,Park2010LaminarSpeeds,Hinman2017ReynoldsWakes}, the base pressure ratio is consistently observed to be $p_\mathrm{B}/p_{02} = 0.03\pm0.01$, where $p_\mathrm{B}$ denotes the pressure at the base of the cylinder (denoted as `B' in Fig.~\ref{fig:cylWake_schem}), and $p_{02}$ represents the pressure at the forward stagnation point. Measurements from recent experiments at Mach 6 show a pressure ratio $p_\mathrm{B}/p_{02} = 0.025$ \cite{Thasu2024thesis}, which is in good agreement with earlier literature. Since significant pressure gradients are not expected in the recirculation region, the pressure downstream of the separation shock wave ($p_\mathrm{i}$) is taken to be $p_\mathrm{i} = p_\mathrm{B}$. By using the flow properties at the pressure minimum location as the upstream conditions and $p_\mathrm{i}$ as the downstream condition, a solution for the separation shock wave angle ($\beta$) and strength can be obtained using standard oblique shock wave relations \cite{Liepmann1957ElementsDynamics}.

In the recirculation region the flow velocities are relatively very low, and hence the region is regarded as stagnant. The acoustic speed in this region, denoted as $a_\mathrm{r}$, is estimated by determining the average local temperature $T_\mathrm{r}$ within the recirculation region. The recovery temperature is a good estimate for $T_\mathrm{r}$ since it accounts for viscous losses in the shear layer \cite{Anderson1984FundamentalsAerodynamics,kumar2024model}. The recovery temperature is given by
\begin{equation}\label{eq:temp_recirc}
    T_\mathrm{r} = \left[\frac{1+\sqrt{Pr}\left(\frac{\gamma-1}{2}\right)M_\mathrm{i}^2}{1+\left(\frac{\gamma-1}{2}\right)M_\mathrm{i}^2}\right]T_0\,,
\end{equation}
where $T_0$ is the stagnation temperature, $Pr$ is the Prandtl number, $\gamma$ is the ratio of specific heats for the fluid, and $M_\mathrm{i}$ is the Mach number downstream of the separation shock wave. With $R$ as the gas constant, $a_\mathrm{r}$ is then written as
\begin{equation}\label{eq:a_recirc}
    a_\mathrm{r} = \sqrt{\gamma R T_\mathrm{r}}\,.
\end{equation}

The propagation speed of the vortical disturbances ($kU_\mathrm{i}$) is estimated by modeling the compressible shear layer as a 2D mixing layer, with supersonic flow on the top and stagnant fluid on the bottom. Extensive literature is available on compressible mixing layers, including studies on growth rate of shear layer thickness and the convection speed of disturbances \cite{Bogdanoff1983Compressibility,Papamoschou1988Compressible,Elliot1990Compressibility,Hall1993Experiments,murray2001characteristics,Pantano2002Study}. Flow instabilities in mixing layers consist of three families of waves, labeled as ``Kelvin-Helmholtz,'' ``supersonic,'' and ``subsonic'' instability waves \cite{Tam1989OnJets,Oertel1979MachJets,Oertel1983CoherentJet}. Propagation speeds of these waves can formally be obtained through linear stability analysis of the mixing layer \cite{Tam1989OnJets}. For the present purpose, however, a simple vortex train model of these instability waves \cite{OertelSen2016MachInteraction} is used to obtain reasonably accurate estimates for the propagation speeds. The vortex train model gives the propagation (or convection) speeds $w_\mathrm{sup}$, $w_\mathrm{KH}$, $w_\mathrm{sub}$ of supersonic, Kelvin-Helmholtz, subsonic instability waves, respectively, as
\begin{equation}\label{eq:oertel}
\begin{split}
\frac{w_\mathrm{sup}}{U_\mathrm{i}} = \frac{1}{1+\alpha} &\\
\frac{w_\mathrm{KH}}{U_\mathrm{i}} = \frac{1+\alpha\left(\frac{w_\mathrm{sup}}{U_\mathrm{i}}\right)}{1+\alpha} = &\; \frac{1+2\alpha}{\left(1+\alpha\right)^2}\\
\frac{w_\mathrm{sub}}{U_\mathrm{i}} = \frac{1-\alpha\left(\frac{w_\mathrm{sup}}{U_\mathrm{i}}\right)}{1+\alpha} = &\; \frac{1}{\left(1+\alpha\right)^2}.
\end{split}
\end{equation}
Here $\alpha = a_\mathrm{i}/a_\mathrm{r}$ is the ratio of sound speeds between the supersonic flow side and the stagnant flow side of the shear layer (see Fig.~\ref{fig:cylWake_schem}). It is noted that all three non-dimensional propagation speeds ($w_\mathrm{sup}/U_\mathrm{i}$, $w_\mathrm{KH}/U_\mathrm{i}$, $w_\mathrm{sub}/U_\mathrm{i}$) are solely a function of $\alpha$.

Based on harmonic analysis of linearized governing equations of compressible inviscid mixing layer flow \cite{Tam1989OnJets}, some key observations of the instability wave characteristics are made here. At low supersonic Mach numbers, only the Kelvin-Helmholtz and subsonic instability waves are active, with the Kelvin-Helmholtz waves dominating the flow. Supersonic instability waves emerge only when $M_\mathrm{i} > 1 + (1/\alpha)$ \cite{Tam1989OnJets}. The growth rates of Kelvin-Helmholtz and supersonic instability waves depend on $M_\mathrm{i}$ and $\alpha$. Specifically, as $M_\mathrm{i}$ increases, the dominance of Kelvin-Helmholtz instability waves decreases while the growth rate of supersonic instability waves steadily rises. The Mach number $M_\mathrm{i}$ above which supersonic instabilities become dominant is termed the critical Mach number. Subsonic waves are active only when the mixing layer has a finite (but small) thickness. Their growth rates are small, and hence they are considered to be the least unstable of the three wave families \cite{Tam1989OnJets}.
\begin{figure}
    \centering
    \includegraphics[width=0.48\textwidth]{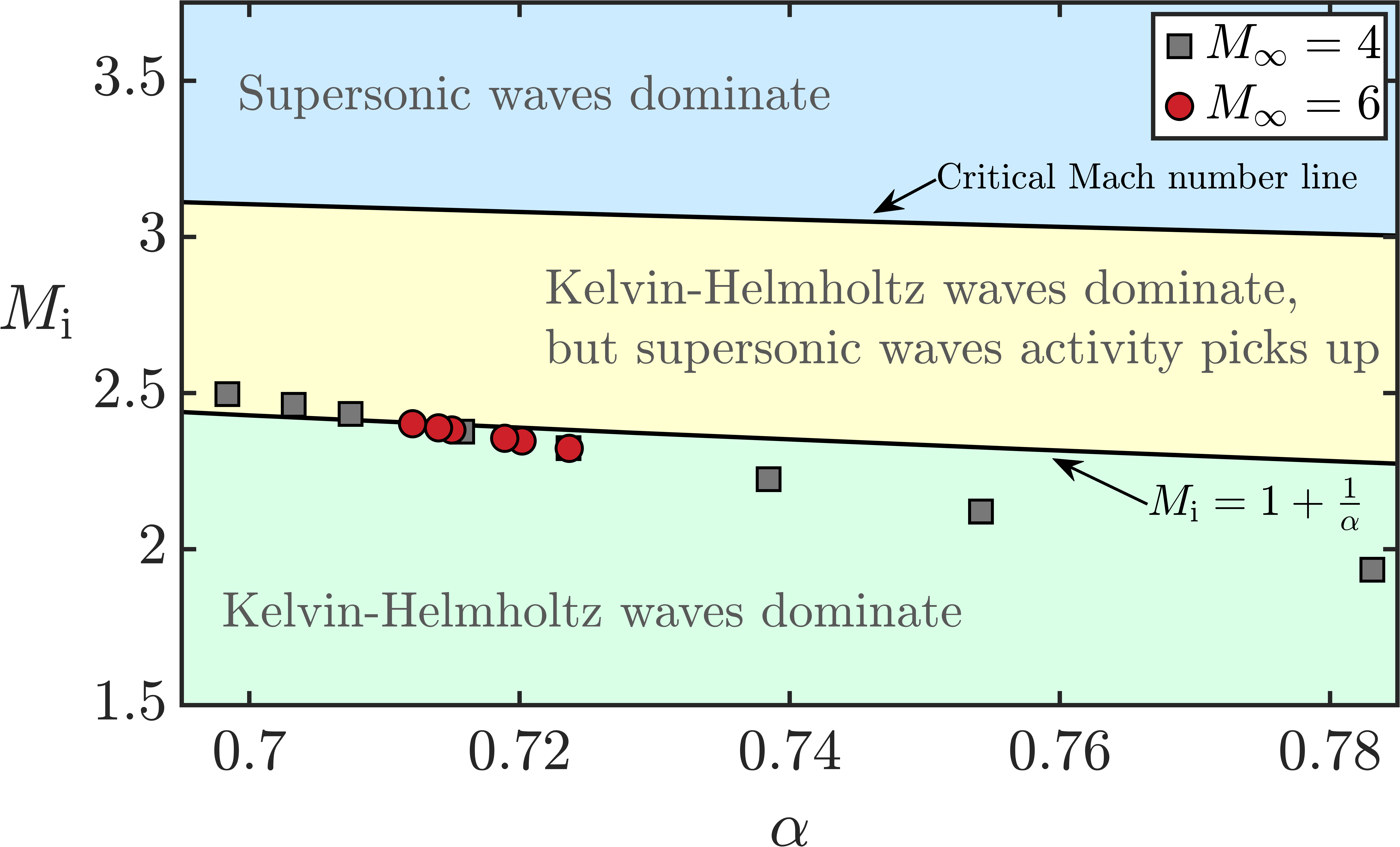}
    \caption{A map of dominant instability waves in the [$M_\mathrm{i}$, $\alpha$] parameter space. Square and circle markers correspond to experimental data given in Table~\ref{tab:exp_data}.}
    \label{fig:DomMap}
\end{figure}
Fig.~\ref{fig:DomMap} shows a map of the [$M_\mathrm{i}$, $\alpha$] parameter space, wherein the $M_\mathrm{i}$ and $\alpha$ values for the experimental data points given in Table~\ref{tab:exp_data} are estimated using the modeling approach outlined earlier in this paper. The figure clearly shows that Kelvin-Helmholtz instability waves are expected to be the dominant instability waves in the wake shear layers across all the experimental data points considered here. Hence, the propagation velocity of vortical disturbances ($kU_\mathrm{i}$) in Eqns.~\eqref{eq:ross_original} and \eqref{eq:ross_StD} is taken to be the convection velocity of the Kelvin-Helmholtz instability waves $w_\mathrm{KH}$ (Eq.~\ref{eq:oertel}); we have
\begin{equation}\label{eq:vort_speed}
kU_\mathrm{i} = w_\mathrm{KH} = \frac{1+2\alpha}{\left(1+\alpha\right)^2}\,U_\mathrm{i}.
\end{equation}

\begin{figure}
    \centering
    \includegraphics[width=0.48\textwidth]{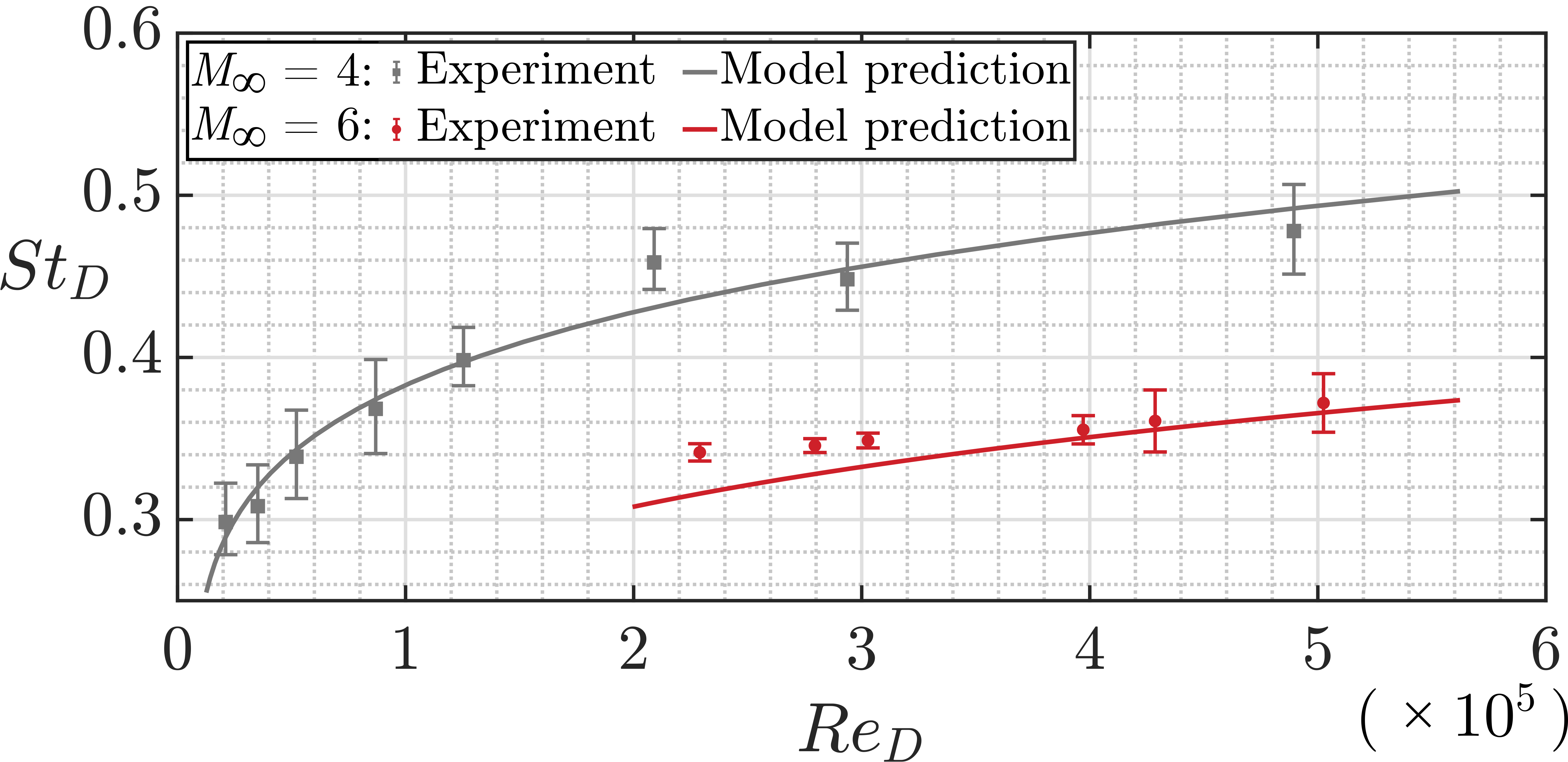}
    \caption{Strouhal number comparison between experimental data and predictions from present model.}
    \label{fig:StD_comp}
\end{figure}

The shear layer length $S$ for use in Eq.~\eqref{eq:ross_StD} is obtained from experimental data given in Table~\ref{tab:exp_data}. The mode number in Eq.~\eqref{eq:ross_StD} is taken as $m = 2$ \footnote{This choice is guided by frequencies observed in experiments \cite{Schmidt2015OscillationsMach4, Thasu2022Strouhal}, which suggest that the second mode frequency is active in the flow}. Based on open cavity flow observations \cite{Rockwell1979Self-sustainedLayers}, the value of $\phi$ is set to 0.25, which corresponds to a phase difference of $90^{\circ}$ between the acoustic and the vortical disturbances at the neck. With that, the modeling exercise is complete, and $St_D$ can be predicted using Eq.~\eqref{eq:ross_StD}. The Strouhal number predictions from the model are compared against experimental data in  Fig.~\ref{fig:StD_comp}. The model is seen to perform well in predicting the experimental measurements at both $M_{\infty} = 4$ and 6. Hence, this exercise lends clear support to the hypothesis that the aeroacoustic mechanism outlined here drives the near-wake oscillations.

\begin{figure}
    \centering
    \includegraphics[width=0.48\textwidth]{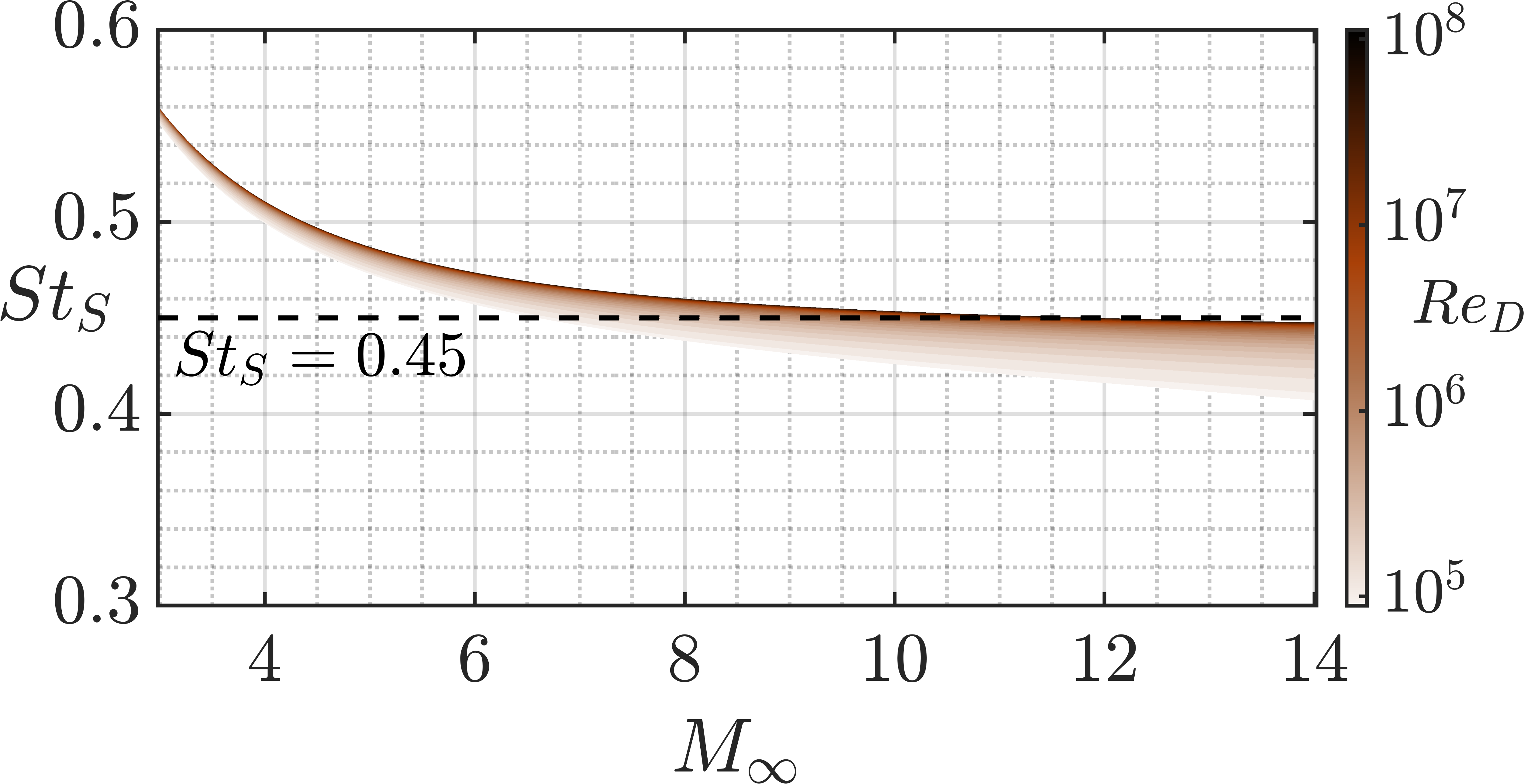}
    \caption{Model prediction for Strouhal number variation with the two governing parameters, $M_\infty$ and $Re_D$.}
    \label{fig:StS_model}
\end{figure}

We now consider $St_S$:
\begin{equation}\label{eq:ross_StS}
    St_S = St_D \left(\frac{S}{D}\right) = \frac{m-\phi}{\frac{U_\infty}{a_\mathrm{r}}+\frac{U_\infty}{kU_\mathrm{i}}}.
\end{equation}
Unlike $St_D$, $St_S$ does not depend on the geometric parameters $S$, $D$. The numerator $\left(m-\phi\right)$ in the above equation is a constant, and the first and the second terms of the denominator account for the roles of acoustic waves and vortical disturbances, respectively, in the feedback loop. From Eqs.~\eqref{eq:temp_recirc}, \eqref{eq:a_recirc}, \eqref{eq:vort_speed}, it is seen that the denominator in Eq.~\ref{eq:ross_StS} depends only on the flow conditions downstream of the separation shock wave (region i) and $\alpha$, both of which in turn depend on $M_\infty$ and $Re_D$. Hence we write 
\begin{equation}
    St_S = g\left(M_\infty,Re_D\right),
\end{equation}
where $g$ indicates functional dependence. The above equation essentially reiterates the fact that the Mach and Reynolds numbers are the two governing parameters for the cylinder wake in the compressible flow regime. The function $g$ constructed using the aeroacoustic model (\emph{i.e.}, by using Eq.~\ref{eq:ross_StS}) is shown in Fig.~\ref{fig:StS_model} \footnote{For the entire range of $M_\infty$ and $Re_D$ considered here, $M_\textrm{i}$ and $\alpha$ are found to be such that Kelvin-Helmholtz waves are dominant throughout. Hence Eq.~\eqref{eq:vort_speed} is used without any modifications for generating Fig.~\ref{fig:StS_model}.}.
$St_S$ shows a considerable dependence on $M_\infty$ at low values of $M_\infty$, whereas the dependence becomes increasingly weaker at higher $M_\infty$. $Re_D$ has a relatively smaller effect, with $St_S$ showing only a small variation across four orders of $Re_D$. The model predicts that $St_S$ attains an invariant value of 0.45 at large $M_\infty$ and $Re_D$. Considering the simple nature of the model \footnote{The model neglects the effects of viscosity and non-isentropic behavior of flow around the cylinder forebody.}, the prediction is found to be very close to the value of 0.48 reported in literature \cite{Schmidt2015OscillationsMach4, Thasu2022Strouhal}. Further, the model predicts that the universal behavior breaks at lower supersonic Mach numbers, where $St_S$ is expected to be sensitive to $M_\infty$.

\begin{figure}
    \centering
    \includegraphics[width=0.48\textwidth]{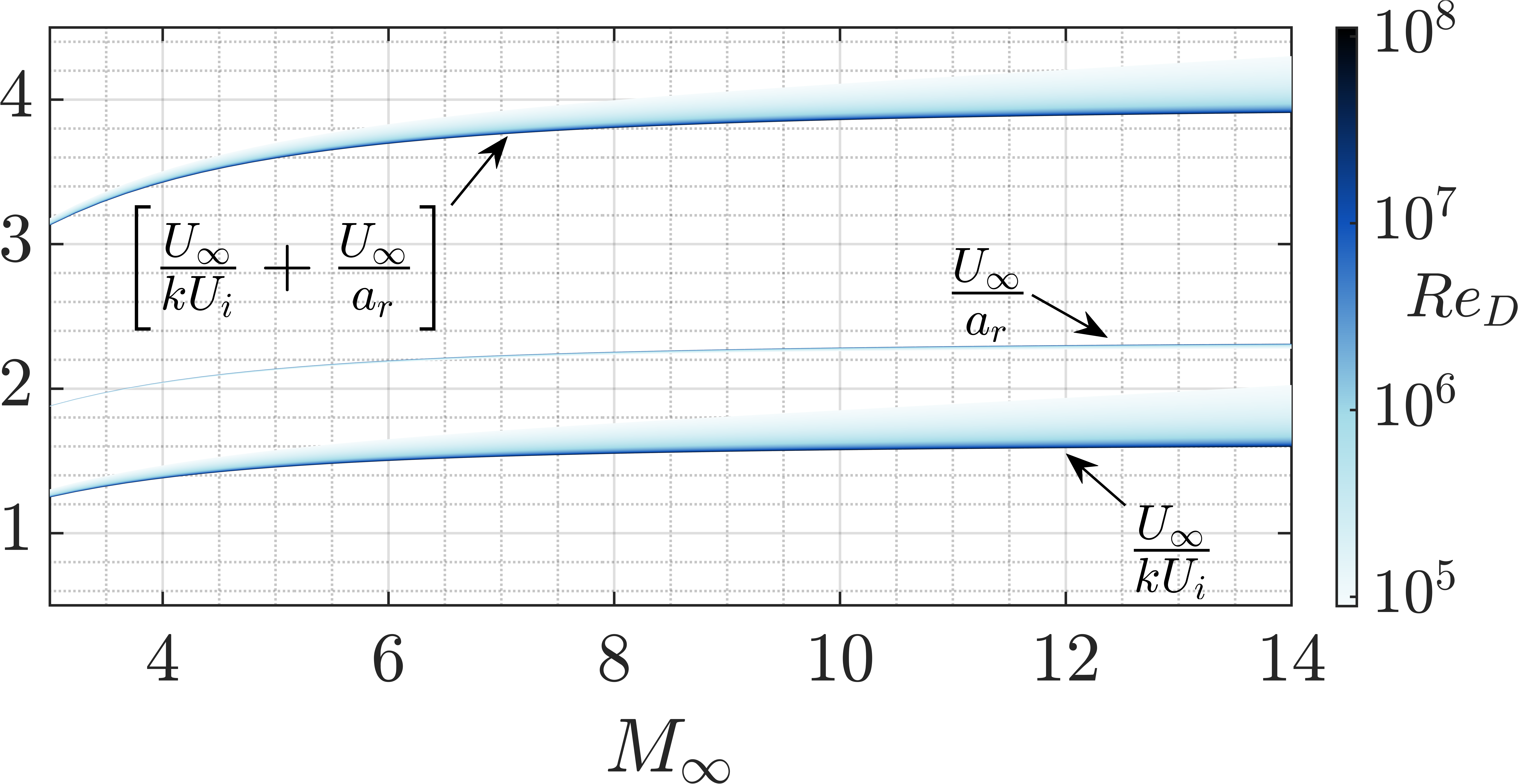}
    \caption{Variation in propagation speeds of vortical and acoustic disturbances.}
    \label{fig:speed_variation}
\end{figure}
It is noted that the vortical and acoustic disturbance propagation speeds ($kU_\mathrm{i}$ and $a_\mathrm{r}$, respectively) are the velocity scales relevant for the feedback loop. We now consider behavior of these disturbance propagation speeds when scaled by the freestream flow velocity, \emph{i.e.}, the quantities $(kU_\mathrm{i}/U_\infty)$ and $({a_\mathrm{r}}/U_\infty)$. Fig.~\ref{fig:speed_variation} shows the variation of $(kU_\mathrm{i}/U_\infty)^{-1}$ and $({a_\mathrm{r}}/U_\infty)^{-1}$ with $M_\infty$ and $Re_D$ obtained from the model \footnote{Variation is shown for inverse of the scaled propagation speeds since the denominator of Eq.~\ref{eq:ross_StS} contains inverse of the speeds.}. It is seen that at high Mach and Reynolds numbers, the disturbance propagation speeds scaled by $U_\infty$ become invariant (and thereby the quantity $[U_\infty/a_\mathrm{r} \,+\, U_\infty/kU_\mathrm{i}]$ in Eq.~\ref{eq:ross_StS} also becomes invariant). Therefore, $U_\infty$ can be treated as the single relevant velocity scale for the feedback loop. Further, it is noted that the length over which both vortical and acoustic disturbances propagate is $S$, which naturally makes it the relevant length scale for oscillations. From these arguments we conclude that $U_\infty$ and $S$ are the appropriate velocity and length scales, respectively, to form the Strouhal number (\emph{i.e.}, $St_S$). And, when the oscillation frequency $f$ is scaled with $S$ and $U_\infty$, we should expect to see invariant behavior at high Mach and Reynolds numbers. This explains the universal behavior of $St_S$ observed in experiments. 


\bibliography{cyl_ref}

\end{document}